\documentclass[11pt,draftclsnofoot,onecolumn]{IEEEtran}
\usepackage{float}
\usepackage[mathscr]{eucal}
\usepackage{ifpdf}
\usepackage{cite}
\usepackage{graphicx}
\usepackage[cmex10]{amsmath}
\usepackage{amssymb}
\usepackage{algorithmic}
\usepackage{units}
\usepackage{setspace}
\usepackage{algorithm}
\usepackage{array}
\usepackage{amsthm}
\usepackage{diagbox}
\usepackage{multirow}
\usepackage{enumerate}
\usepackage{color}
\usepackage{xcolor}
\usepackage{mathtools}
\newcommand{\subparagraph}{}
\usepackage[compact]{titlesec}
\usepackage{lmodern}
\usepackage{graphics}
\usepackage{chngcntr}
\usepackage{booktabs}
\usepackage{multirow}
\usepackage{float}
\usepackage{tabularx}
\usepackage{soul}
\usepackage[normalem]{ulem}
\usepackage[utf8]{inputenc}

\colorlet{customBlue}{blue!30}
\colorlet{customYellow}{yellow!30}
\colorlet{customRed}{red!30}

\begin{document}
\title{Personalized Education in the AI Era:\\
What to Expect Next?}
\author{
\IEEEauthorblockN{Setareh Maghsudi, Andrew Lan, Jie Xu, and Mihaela van der Schaar}
\thanks{S. Maghsudi is with the Department of Computer Science, University of Tübingen, Germany (email: setareh.maghsudi@uni-tuebingen.de). A. Lan is with the College of Information and Computer Sciences, University of Massachusetts Amherst, MA, USA (email: andrewlan@cs.umass.edu). J. Xu is with the Department of Electrical and Computer Engineering, University of Miami, FL, USA (email: jiexu@miami.edu). M. van der Schaar is with the Faculty of Mathematics, University of Cambridge, UK (email: mv472@damtp.cam.ac.uk). }
}
\maketitle
\begin{abstract}
The objective of personalized learning is to design an effective knowledge acquisition track that matches the learner’s strengths and bypasses her weaknesses to ultimately meet her desired goal. This concept emerged several years ago and is being adopted by a rapidly-growing number of educational institutions around the globe. In recent years, the boost of artificial intelligence (AI) and machine learning (ML), together with the advances in big data analysis, has unfolded novel perspectives to enhance personalized education in numerous dimensions. By taking advantage of AI/ML methods, the educational platform precisely acquires the student’s characteristics. This is done, in part, by observing the past experiences as well as analyzing the available big data through exploring the learners’ features and similarities. It can, for example, recommend the most appropriate content among numerous accessible ones, advise a well-designed long-term curriculum, connect appropriate learners by suggestion, accurate performance evaluation, and the like. Still, several aspects of AI-based personalized education remain unexplored. These include, among others, compensating for the adverse effects of the absence of peers, creating and maintaining motivations for learning, increasing the diversity, removing the biases induced by the data and algorithms, and the like. In this paper, while providing a brief review of state-of-the-art research, we investigate the challenges of AI/ML-based personalized education, and discuss potential solutions. 
\end{abstract}
\section{Introduction}
\label{sec:COPE}
\begin{figure*}[tp]
    \centering
    \includegraphics[width=.80\columnwidth]{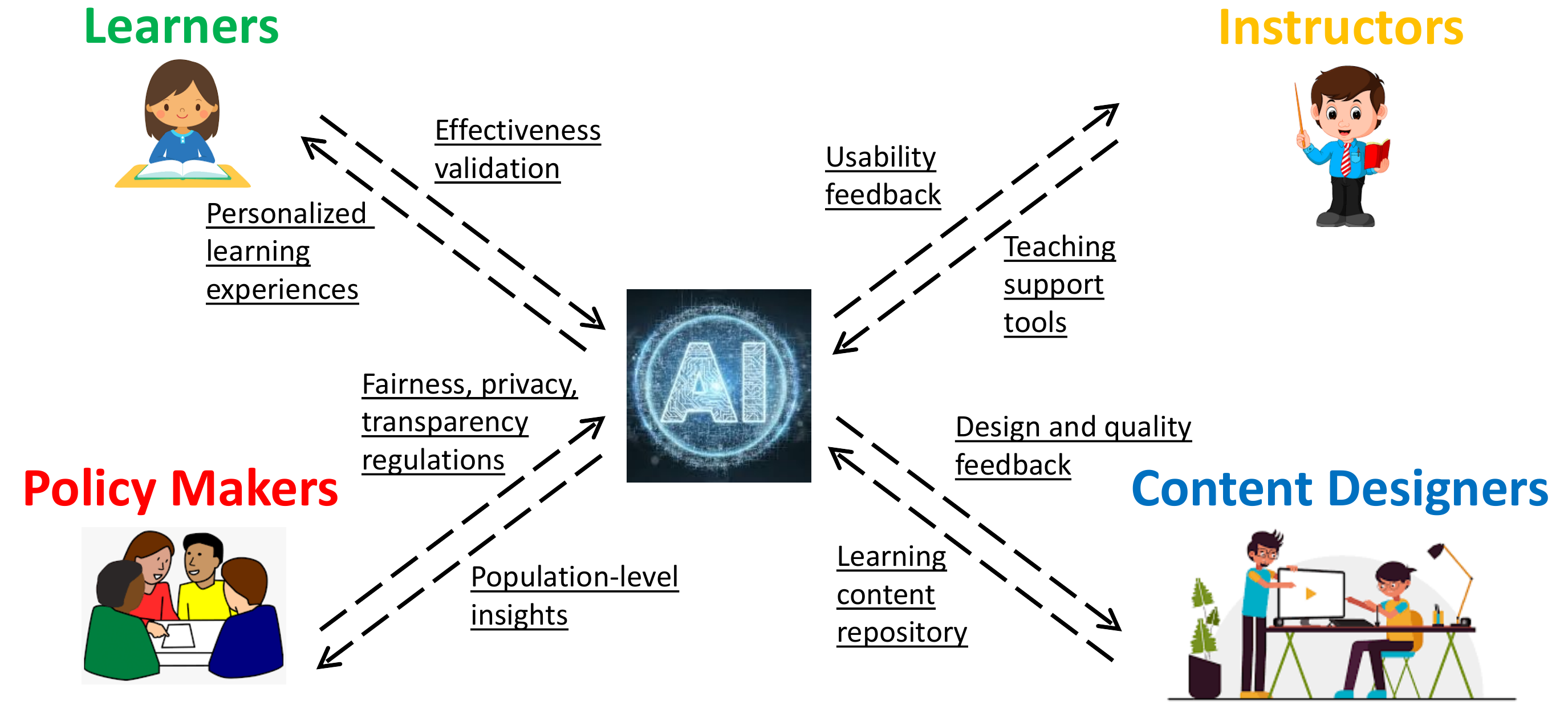}
    \caption{The baseline ecosystem of AI-empowered personalized education.}
    \label{fig:Ecosys}
\end{figure*}

The last decade has witnessed an explosion in the number of web-based learning systems due to the increasing demand in higher-level education, the limited number of teaching personnel, advances in information technology and artificial intelligence, and, more recently, COVID-19. In the past few years, to enhance the conventional classrooms, to bridge the constraints of time and distances, and to improve fairness by making high-quality education accessible, most universities have integrated Massive Open Online Course (MOOC) platforms such as the edX consortium in their education systems. Also, several schools have added online labs to their structures, where students, especially those who cannot access physical labs, can perform experiments. Besides, there has been significant growth in the development of other online educational tools that simplify learning. These include, for example, the software for text summarization in different domains, also to produce questions and tests, followed by evaluation, which can be great assistance not only to students but also to teachers. Several advantages of these systems over traditional classroom teaching are: (i) They provide flexibility to the student in choosing what to learn and when to learn; (ii) They do not require the presence of an interactive human teacher; (iii) There are no limitations in terms of the number of students who can participate in the course. \textbf{Figure \ref{fig:Ecosys}} shows the baseline ecosystem of online personalized education, including all the stakeholders, together with the crucial factors and performance metrics.

However, the currently available online teaching platforms have significant limitations. To a large extent, personalized education has been mainly diminished to a specific type of 'recommender system', although its potential goes far beyond advising a series of lectures on an online platform that might be interesting to a specific user. One fundamental difference between existing recommender systems and personalized education is the optimization objective: The former focuses on some form of user engagement to maximize profit, which is system-centric and relatively easy to quantify, whereas the latter focuses on some form of learning outcomes, which is student-centric and hard to define.

\begin{figure*}[tp]
    \centering
    \includegraphics[width=.98\columnwidth]{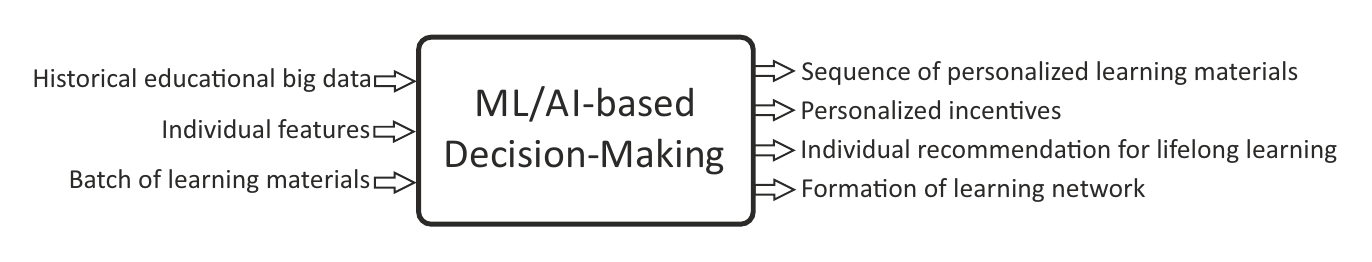}
    \caption{The basic concept of AI-powered personalized education.}
    \label{fig:Basic}
\end{figure*}

ML/AI-enabled education is a response of great potential to the current shortcomings. It creates a new and more flexible learning technology genre that adapts to student learning and allocates resources as obliged. It takes advantage of the strengths of both online tools and individual tutoring. As such, AI-enabled personalized education promises to yield many of the benefits of one-on-one instruction at a per-student cost similar to large university lecture classes. The system applies to both online courses and courses that have a hybrid of classroom and on-line instruction. As shown in \textbf{Figure \ref{fig:Basic}}, ML/AI-enabled education comprises of a large set of decision-making strategies that collectively map the available data together with the individual features to a variety of personalized educational materials and recommendations. 

\begin{figure*}[tp]
    \centering
    \includegraphics[width=.5\columnwidth]{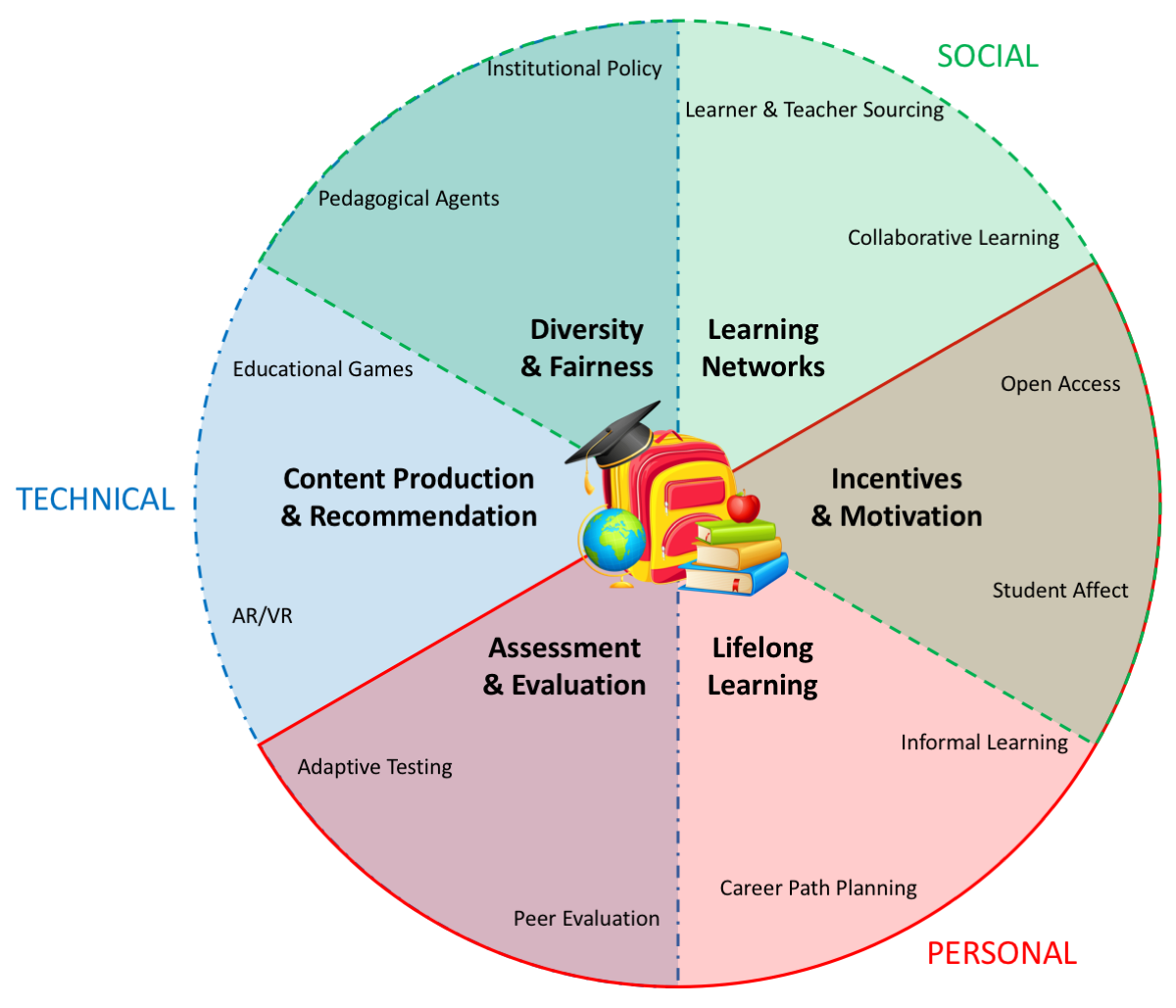}
    \caption{A list of (some of) the topics in personalized education, organized by three different aspects: technical, personal, and social. We focus on six of them in this paper.}
    \label{fig:overview}
\end{figure*}

Data can be collected on performance in both traditional assignments (problem sets, computer programs, laboratory) as well as online exercises and tests. It includes built-in assessment tools as an essential part of its optimization of lesson sequences. As such, it supports the educational community in developing new teaching modalities in a broad range of disciplines. However, despite intensive research efforts in this decade, a variety of aspects of personalized education remain unexplored, including both dark- and bright sides. In this paper, we discuss six core topics, review existing work, outline their limitations, and propose future research directions; see \textbf{Figure~\ref{fig:overview}} for an overview. 

When discussing any form of education, \textit{quality} is an inevitable keyword. The quality of education depends largely on the quality of the available learning content and the quality of the personalized recommendations that guide each learner to the most suitable learning content. So far, the researchers have studied the production of learning content, from developing AI-driven smart learning content such as intelligent, interactive textbooks and game-based learning platforms to automatically generating learning content from the wild. Reference \cite{weixu10:DLF}, for example, develops a sentence deletion method for text simplification. Besides, in \cite{sachan}, the authors investigate the effectiveness of discourse in multimedia to extract the knowledge from the textbooks. Moreover, a large body of the existing work investigates the recommendation of both macro-level and micro-level learning content, including courses in learners' degree plans as well as specific remedial content such as lecture notes, videos, and practice problems. For example, in \cite{Manickam17:CMAB}, the authors take advantage of a multi-armed bandit framework to optimize the selection of learning resources and questions to satisfy the needs of each individual student. Moreover, Reference \cite{Ghauth10:LMR} develops an e-learning recommender system framework based on two concepts: peer learning and social learning, which encourage students to cooperate and learn jointly. Despite great efforts, there remain several challenges to address. These include content recommendation at heterogeneous levels, the recommendation of a bundle of connected contents followed by performance evaluation, and the Pareto-optimization of conflicting objectives in the content recommendation. We discuss these progress and the future steps in \textbf{Section \ref{sec:CPR}}.

Historically, education is tightly coupled with evaluation. In personalized education, assessment and evaluation concern both the learner's performance and the effectiveness of the intelligent learning platform. Early approaches for learner's assessment such as the 'classical testing theory' (CTT) use summaries of graded standardized tests. Recent approaches include 'item response theory' (IRT) models that enable the estimation of latent knowledge mastery levels and knowledge tracing models that pursue the evolution of a learner's knowledge. In \cite{Magno09:DDC}, the authors compare the CTT and IRT. Methods such as 'computerized adaptive testing' improve the efficiency of assessments. The current approaches to evaluate the learning platforms use rigorous experiments, often large-scale randomized controlled trials. In this area, open problems include the prediction of learners' future performance, which enables providing better recommendations and more accurate feedback. This is referred to as the knowledge tracing (KT) problem, for which several methods are developed in the past few decades. As an example among many others, \cite{pardoskt} discusses a Bayesian framework for KT. Another challenge is to reduce the information loss while grading the arrived input from the learner, by accurate interpretation of knowledge level based on the test design. We elaborate and address such challenges in \textbf{Section \ref{sec:AESP}}.

The huge advances in science, technology, and healthcare have changed the working life of humans. Individuals have way more alternatives to choose a job, they tend to change their job more frequently than before, are more open to mobility, and the career spans a long period of life. As such, continuing education, which aims at advancing one's educational process, as well as lifelong learning, i.e., pursuing additional professional qualifications, are important components of educational policy in the world. Implementing these two concepts successfully has a significant impact on social welfare by developing new skills that enhance personal- and professional life. During the past decade, AI-/ML-based personalized education has been under intensive investigation from several perspectives; nonetheless, the aforementioned aspects are largely neglected. Indeed, personalized education shall accompany the learner throughout her life, which can be difficult and costly to implement. Other challenges include lack of appropriate data, potentially long delay to feedback, high diversity, and fast dynamics in the environment. For example, in \cite{Sharples00:DPM}, the authors design a new genre of educational technology-personal computer systems- that support learning from any location throughout a lifetime. Another research direction is to enable learning system to learn continuously. Reference \cite{Parisi19:CLL}, e.g., investigates the ability of neural networks to enable life-long learning. We elaborate more on this topic in \textbf{Section \ref{sec:LLL}}.

Similar to any other task, humans require motivation for learning. Generally, incentives for learning can be defined as an inducement or supplemental reward that serves as a motivational device for intended learning \cite{Grove12:ISL}. Presumably, the most conventional models of incentive are the 'grade' and 'certificate', which are implemented as a part of learning platforms to motivate the students. The strength of such motivation depends on the validity and acceptance of such certificates by different authorities such as employers. Nonetheless, employing AI methods enables for incentive design far beyond handing a certificate. This includes, for example, monetary rewards in the form of bonuses for online learning materials. The incentives can be induced also by soft-methods such as gamification based on the learner's characters to promote continuous learning, or adapting the features of the learning environment based on the learner's traits to engage her in the learning process as far as possible. In \cite{Buckley14:ILE}, for example, the authors investigate the effects of gamification on students' motivation from several perspectives. Besides, \cite{Chen10:AAA} discusses several factors on motivation in online learning together with their relative salience. We discuss such challenges and methods in \textbf{Section \ref{sec:IM}}.

Education is social and learners can extremely benefit from their peers. Therefore, it is urgent to develop effective ways to build networks that serve as a conduit of knowledge for learners to interact with each other. In the current form, personalized education suffers from a lack of student-student and student-teacher connections and interactions, which have an unquestionably positive impact on learning through discussions, joint efforts, and brainstorming. In \cite{Rovai00:BSC}, the authors study building and sustaining community in asynchronous learning networks, i.e., when the learners are physically separated. Moreover, Reference \cite{Vesely07:KEB} investigates and compares the influence of such communities from both students'- and teachers' perspectives. Despite the past research efforts, we believe that capitalizing on AI and ML methods, online platforms have more to offer, especially, for building the knowledge- and expertise networks that facilitate the assimilation and dissemination of knowledge, and consequently, by enabling close interactions (in terms of mentorship, friendship, coworkers, and the like), creating knowledge. Personalized education platforms can promote autonomous network formation by encouraging learners to interact. Moreover, the platforms can establish links among those learners that satisfy some similarity conditions and hence can be useful to each other for cooperation, inspiration, and motivation. We elaborate on these issues in \textbf{Section \ref{sec:BLE}}.
 
In many different ways, education affects the well-being of humans, and thereby the society, both in the short-term and long-term. As such, fairness is a highly important aspect of education, regardless of being in conventional classrooms form or in modern platforms that can personalize the learning experience. Despite this great importance, personalized education, similar to its traditional counterpart, might result in- and strengthen inequality. This arises, for example, due to unequal access to the learning platforms, biases in training data, inaccuracy in algorithm design, and the like. Indeed, existing research shows that some subgroups of students, mainly those privileged also in conventional education forms, would profit from personalized education more than their peers. To address such issue more religiously, there has been intensive effort to develop appropriate fairness models \cite{Gardner:Evaluating:2019}. Moreover, several research works such as \cite{Yao:Beyond:2017} study the fairness of predictive algorithms in educational settings. Another crucial issue is 'diversity'. Today, it is well-established that diversity promotes innovation and efficiency in the working place. Nonetheless, given the social-responsibility of education, only recruiting diverse talents does not suffice. AI-based personalized education platform can be a boost to diversify the education environment, for example, by rewarding collaborative learning in diverse networks. We discuss these topics in \textbf{Section \ref{sec:DFB}}.
\section{Content Production and Recommendation}
\label{sec:CPR}
The quality of education ultimately depends on the quality of the learning content. Creating new content requires the wisdom of human content designers and educational experts; to date, AIs have not shown the capability of creating learning content on their own. However, they still have plenty to offer in content production by automating mundane jobs and helping humans in tasks where human input is necessary. Specifically, the role of AI should be to i) take away repetitive tasks that can be automated and ii) assist humans by providing feedback extracted from data during the process of content production in a human-in-the-loop manner. There are ample future research directions in content production; we list a few below. 
\begin{itemize}
\item \textbf{Content summarization and question generation:} In many educational domains, knowledge is factual. For example, in History, one often needs to remember specific detailed facts about historical events. Even in scientific domains such as Biology, there is also factual knowledge such as the size and life span of an animal. In this case, there are many natural language processing (NLP)-based tools that can be used for content production. For example, text summarization tools can sort through long, sometimes redundant textbook sections and extract key facts for remedial studies. This is not only helpful but also sometimes crucial to certain learner groups, such as those with learning disabilities. Moreover, automatic question generation can effectively produce high-quality factual assessment questions that have short, textual answers \cite{qgen}. An example of automated question generation is shown in \textbf{Table~\ref{tbl:qgen}}; we reversed a long short-term memory (LSTM) network-driven question answering pipeline trained on common question answering datasets, turned it into a question generation pipeline, and applied it to textbooks. Human experts have indicated that the quality of generated questions is higher than that generated by other methods. See \cite{qgen} for details.
\item \textbf{Multi-modal content understanding:} Many educational domains involve multi-modal learning content, such as text,  formulas, figures, and diagrams. When a learner fails to answer an assessment question correctly, personalized education systems need to automatically retrieve relevant content to help the learner resolve their confusion (by retrieving examples and explanations) or give the learner more practice opportunities (by retrieving assessment questions). Retrieving content within the same modal is relatively easy; for example, when a learner answers a textual question incorrectly, it is possible to use information retrieval methods to extract relevant textbook chunks or lecture slides. However, when the most helpful content is in another modality, for example, when a Venn diagram is the most effective at helping a learner clear up a misconception in a probability question involving text and mathematical formulas, it is hard to retrieve the diagram. Therefore, more work needs to be done when the domain includes multi-modal content; To understand these content modalities and use them for content production, we need to learn universal representations across all modalities, possibly using embedding approaches to map multiple modalities into a shared vector space \cite{lafferty}.
\item \textbf{Human-in-the-loop content design:} Even for humans, learning content is not created in one shot; just like textbooks have different editions, learning content is frequently edited and updated over time. Therefore, during this multi-step process, we can use AIs to act as (possibly even interactive) assistants to content designers. Duties that can be assigned to AIs include i) Analyzing learners' data to identify the areas of priority for new content and assessment questions that need to be improved (see Section~\ref{sec:AESP} for discussions on how existing learner assessment models can also provide information on question quality); ii) Providing drafts of instructor responses and perform automated checking of human-generated content using NLP tools; iii) Using crowdsourcing to put the learning content together by soliciting on-demand feedback \cite{neilcrowd}. The last task is especially important in online educational settings, where learning follows during frequent exchanges between learners and human instructors and assistants \cite{Zylich:Exploring:2020}.  
\end{itemize}
\begin{table*}[t]
\small
\centering
\resizebox{\textwidth}{!}{
\begin{tabularx}{\textwidth}{XX}
\toprule
\multicolumn{2}{p{\dimexpr\textwidth-2\tabcolsep\relax}}{\textbf{Context (Biology)}: On each chromosome, there are thousands of genes that are responsible for determining the genotype and phenotype of the individual. A gene is defined as a sequence of DNA that codes for a functional product. The human haploid genome contains 
\textcolor{blue}{\hl{{\uline{\textbf{3 billion base pairs}}}}} 
and has 
\textcolor{red}{\hl{{\dashuline{\textbf{between 20,000 and 25,000}}}}} 
functional genes.} \\
\midrule
\multicolumn{1}{c}{\textcolor{blue}{\hl{{\uline{\textbf{3 billion base pairs}}}}}}
& 
\multicolumn{1}{c}{\textcolor{red}{\hl{{\dashuline{\textbf{between 20,000 and 25,000}}}}}}
\\ \cmidrule(lr){1-1} \cmidrule(lr){2-2} 
How many base pairs are on the human genome?
&
How many functional genes are on the human haploid genome? \\
\bottomrule
\end{tabularx}
}
\vspace{.2cm}
\caption{Example of two automatically generated assessment questions for two different answers with the same input context from a textbook. The answers are underlined and marked with different colors in the input context.}
\label{tbl:qgen}
\end{table*}

\begin{figure*}[tp]
    \centering
    \includegraphics[width=.6\columnwidth]{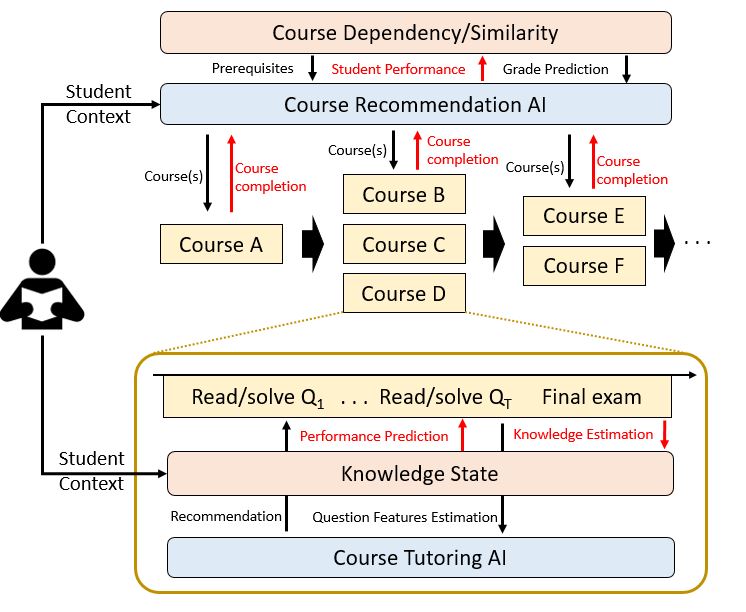}
    \caption{A detailed framework for course recommendation.}
    \label{fig:CourseRecom}
\end{figure*}

Even with high-quality learning content, the presentation, i.e, the personalized recommendation of the right learning content to the right learner at the right time is crucial to optimize the learning outcomes. Fortunately, this is an area where AIs can excel at: By automatically deploying recommendations and analyzing the data of learners' performance, they can quantify the effect of learning content on certain learners in terms of specific learning outcomes to detect the most effective ones. On the contrary, humans, even educational experts in the past, use theoretical models of learning and do not fully take advantage of this data. Among several directions for future research in this area, we discuss a few in the following. 
\begin{itemize}
\item \textbf{Recommendations at the microscopic and macroscopic levels:} Learning content is organized at multiple levels, down to individual paragraphs and assessment questions, and up to courses and textbooks that organize several pieces of learning content together. Therefore, we need to study content recommendations at multiple levels: (i) microscopic level such as individual questions and lecture video suggestions \cite{banditslan}; and (ii) macroscopic level such as course recommendation, especially for learners taking massive open online courses (MOOCs) \cite{xu2016personalized}. 
\item \textbf{Efficient experimentation and synthetic learner models:} Traditionally, the fields of learning science and education have relied on rigorous A/B testing to validate the educational impact of learning content, usually in terms of its ability to improve learning outcomes for learners in the experimental group over those in the control group. However, this approach leads to long experimental cycles since (i) one can typically validate only one learning content at a time, and (ii) metrics such as long-term learning outcomes naturally require long experimental cycles. Therefore, it is imperative to search for novel tools that enable rapid experimentation. Possible solutions include using Bayesian optimization to test multiple contents simultaneously \cite{mozeraz}, or utilizing reinforcement learning (RL) as more and more learners use a piece of learning content. In the past, using RL to learn instructional policies (content recommendation can be viewed as a form of the instructional policy) has been limited due to the lack of large-scale real learner data; however, recent approaches have looked at using data- or cognitive theory-driven synthetic learner models to simulate learner data \cite{pnasrep}.
\item \textbf{Conflicting objectives:} There is no unified objective in personalized learning since learning outcomes itself is defined at multiple timescales. The optimal action may differ across different objectives. For example, the learning content used in a practice session that maximizes a learner's performance on the midterm exam tomorrow may differ from the one that maximizes their overall course grade, which may differ from the one that maximizes their chance of getting a specific job after graduation. Therefore, We need to develop personalization algorithms that can balance multiple objectives and even resolve potential conflicts among these objectives. We also need to understand how these objectives interact with each other; for example, what skills taught in courses and schools carry over after graduation, which is a key issue in lifelong learning (discussed in detail in Section~\ref{sec:LLL}). 
\end{itemize}
\textbf{Figure \ref{fig:CourseRecom}} shows the interplay between different elements, such as context, prediction, feedback, and the like, to optimize the course recommendation. It is worth noting that the approaches described above are generic in the sense that they have wide applicability to different educational areas including signal processing, possibly with minor domain-dependent adaptations. As an example, in \cite{Tekin15:OLP}, the authors apply several of the aforementioned ideas to develop \textit{eTutor}, a personalized web-based education system that learns the optimal sequence of teaching materials to show based on the student's context and feedback about the previously shown teaching materials. In an experiment, they apply the eTutor system in the following scenario: The students have studied digital signal processing in the past. The role of eTutor is to recommend learning materials to the students with the goal of refreshing their minds about discrete Fourier transform in the minimum amount of time. The e-tutor shows better performance compared to a random- and a fixed-selection rule.
\section{Assessment and Evaluation}
\label{sec:AESP}
A key problem in learner assessment is to estimate how well they master each knowledge component/concept/skill from their responses to assessment questions. Related works can be broadly classified into two categories: (i) \emph{static} models that analyze the generated data as learners take an assessment and thereby assuming that each learner's knowledge remains constant during the assessment, and (ii) \emph{dynamic} models that track learners' progress throughout a (possibly long) period as their knowledge levels evolve. Below we provide a short overview of each category. 
\begin{itemize}
\item \textbf{Static models- Item response theory (IRT):} The basic 1PL IRT model characterizes the probability that a learner answers a question correctly as
\begin{align*}
P(y_{i,j}=1)=\sigma(a_j-d_i), 
\end{align*}
where $y_{i,j}$ denotes the binary-valued graded response of learner~$j$ to question~$i$, where $1$ implies a correct response and $0$ otherwise. Moreover, $a_j \in \mathbb{R}$ and $d_i \in \mathbb{R}$ are scalars that correspond to the learner's ability and the question's difficulty, respectively. Also, $\sigma(\cdot)$ is a link function that is usually the sigmoid function or the inverse probit link function \cite{lordirt}. Later extensions include 2PL IRT models that add another multiplicative scaling parameter. This parameter corresponds to the ability of each question differentiating high-capacity learners from low-capacity ones. Besides, 3PL IRT models add another scalar outside of the link function, which corresponds to the probability that an item can be guessed correctly. Finally, multidimensional IRT models use vectors instead of scalars to parametrize the strengths and difficulties to capture multiple aspects of one's ability \cite{mirt}. Using the aforementioned models, one can (i) obtain relatively stable estimates of learners' ability levels by denoising learners' responses and (ii) estimate the quality of each assessment question. 
\item \textbf{Dynamic models- Knowledge tracing (KT):} KT models consist of two parts, learner performance model ($f(\cdot)$) and learner knowledge evolution model ($g(\cdot)$), as
\begin{align*}
y_t \sim f(a_t), \quad a_t \sim g(a_{t-1}),
\end{align*}
where $t$ denotes a discrete-time index. Early KT models such as Bayesian KT \cite{bkt} treats knowledge ($h_t$) as a binary-valued scalar that characterizes whether or not a learner masters the (single) concept covered by a question. The performance and knowledge evolution models are simply noisy binary channels. Later, factor analysis-based KT models use a set of hand-crafted features such as the number of previous attempts, successes, and failures on each concept to represent a learner's knowledge levels \cite{pfa}. These models require expert labels to associate questions to concepts, resulting in excellent interpretability since they can effectively estimate the knowledge level of each learner on expert-defined concepts. Recent KT models incorporate deep learning, especially recurrent neural networks into the KT framework, where knowledge is represented as a latent vector $\mathbf{a}_t$ \cite{dkt}. These models achieve state-of-the-art performance in predicting future learner responses, although in some cases the advantage is not significant despite paying the price of losing some interpretability \cite{howdeep}. 
\end{itemize}

\begin{figure*}[tp]
    \centering
    \includegraphics[width=.7\columnwidth]{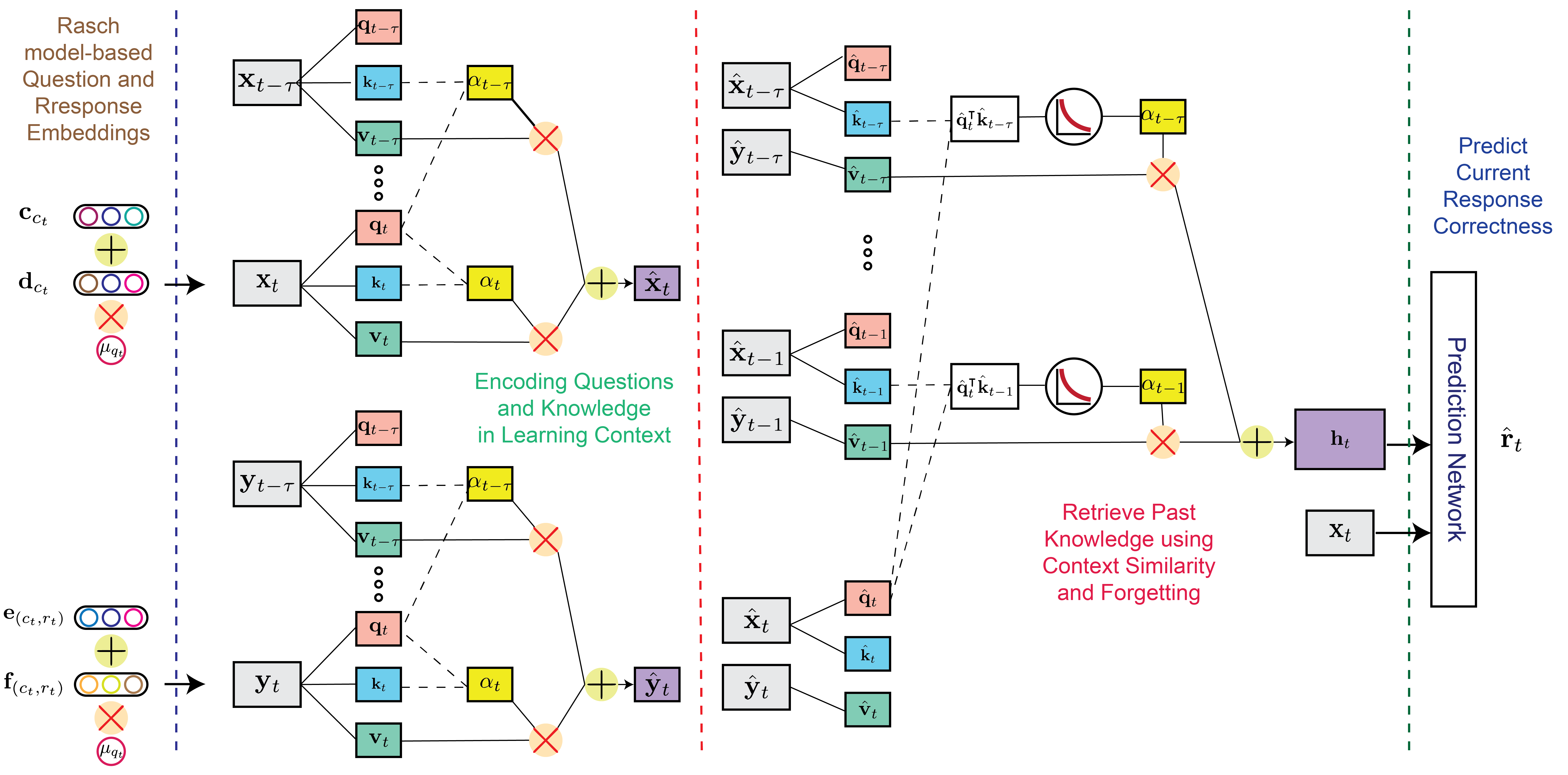}
    \caption{Overview of the AKT method. We use IRT-based raw embeddings for questions and responses. We compute context-aware representations of questions and responses using two encoders. We then use a knowledge retriever to retrieve past acquired knowledge for each learner using a monotonic attention mechanism, which is used for performance prediction.}
    \label{fig:akt}
\end{figure*}

The existing learner assessment models have several bottlenecks. First, there are not many models with both the ability to achieve state-of-the-art performance in data fitting, (i.e., future performance prediction) as well as feedback generation (i.e., providing interpretable feedback to learners and instructors for downstream tasks such as personalization). Therefore, it is imperative to develop new deep learning-based models that not only inherit the flexibility of neural networks to accurately predict learner performance but also build in cognitive theory-inspired structures to promote interpretability and enable the generation of meaningful feedback. As an example, in the recently developed attentive knowledge tracing (AKT) model \cite{akt}, visualized in \textbf{Figure~\ref{fig:akt}}, we combined state-of-the-art attention networks with cognitive theory-inspired modules. We used a monotonic attention mechanism where weights exponentially decay over time and questions embeddings parametrized by the 1PL IRT model to prevent overfitting. Experimental results show that AKT not only outperforms existing KT models but also exhibits some interpretability; see \cite{akt} for details. 
Moreover, existing learner assessment models almost exclusively operate on \emph{graded} learner responses; however, converting raw learner responses to graded responses leads to considerable information loss. For multiple-choice questions, different distractor options are not created equal; choosing certain incorrect options over others might indicate that a learner exhibits a certain misconception. However, this information is lost when the learner's option choice is converted to a graded response. Moreover, due to its superior pedagogical value, open-response questions are widely adopted; the specific open-ended response a learner enters contains rich information about her knowledge state. Therefore, it is vital to develop models that go deeper than the graded response level and into the \emph{raw} response level. These models enable personalization at even finer levels, e.g., after each step as a learner solves an open-ended mathematical problem step-by-step, and enable personalized education systems to attend to learner difficulties in a more timely manner. 

Another consideration in effective learner evaluation is that assessment and performance prediction models must be tailored to different learning environments and platforms. For example, accurate prediction of students' future college performance based on their ongoing academic records is crucial to carry out effective pedagogical interventions so that on-time and satisfactory graduation is ensured. However, foretelling student performance in completing degrees (e.g., college programs) is significantly different from that for in-course assessment and intelligent tutoring systems. In what follows, we describe the most important reasons. 
\begin{itemize}
\item First, students differ tremendously in terms of backgrounds as well as the study domains (majors, specializations), resulting in different selected courses. Even if the courses are similar, the sequences in which the students take the courses might differ significantly. Therefore, a key challenge for training an effective predictor is to handle heterogeneous student data. In contrast, solving problems in intelligent tutoring systems often follow routine steps that are identical for all students. Similarly, predictions of students' performance in courses are often based on in-course assessments that are identical for all students.  
\item Second, although the students often take several courses, not all of them are equally informative for predicting the students' future performance. Utilizing the student's past performance in all courses that he/she has completed not only increases complexity but also introduces noise in the prediction, thereby degrading the performance. For instance, while it is meaningful to consider a student's grade in 'Linear Algebra' for predicting his/her grade in 'Linear Optimization', the student's grade in 'Chemistry Lab' may have much weaker predictive power. However, the course correlation is not always as obvious as in this example. Therefore, to enhance the accuracy of performance predictions, it is essential to discover the underlying correlation among courses.  
\item Third, predicting student performance in a degree program is not a one-time task; rather, it requires continuous tracking and updating as the student finishes new courses over time. An important consideration is the following: The prediction shall be made based on not only the most recent snapshot of the student accomplishments but also the evolution of the student progress, which may contain valuable information to improve the prediction's accuracy. However, the complexity can easily explode since even mathematically representing the evolution of student progress itself can be a daunting task. Treating the past progress equally as the current performance when predicting the future may not be a wise choice either since old information tends to be outdated.
\end{itemize}
Finally, we would like to emphasize the following: Similar to the offline system, in AI-power personalized education, the assessment does not remain limited to evaluating the performance of individual students in different tests in a single online education portal. Indeed, evaluation might be necessary not only for individuals but also for a collection of students as well as other stakeholders such as educators, policy-makers, and the providers of online education. In particular, fair and precise comparison, analysis, and accreditation of online education portals, as well as the degrees and certificates provided by such portals, are crucial. The reasons include the following: (i) Distance education has grown into a broad industry in the past decade; (ii) The majority of the learners rely on certificates of online classes as approval for obtaining the necessary knowledge and skills; (iii) Online education is inherently international and crosses the boundaries. Similar to improving the evaluation of students, AI and ML methods, together with bid data analysis, can assist in accreditation and comparison of online portals and the issued degrees and certificates; this includes, e.g., comparing the average students' performance with an online degree with that of traditional, yet accredited, degree.  A detailed discussion of such topics have several perspectives, and therefore it is out of the scope of this paper.     
\section{Life-long Learning}
\label{sec:LLL}
Life-long learning emphasizes holistic education and the fact that learning takes place on an ongoing basis from our daily interactions with others and with the world around us in different contexts. These include not only schools but also homes and workplaces, among several others. Because of its ongoing nature, making foresighted learning plans is crucial for life-long learning to achieve the desired outcome. 

\begin{figure}[!h]
	\centering
	\includegraphics[width=0.6\textwidth]{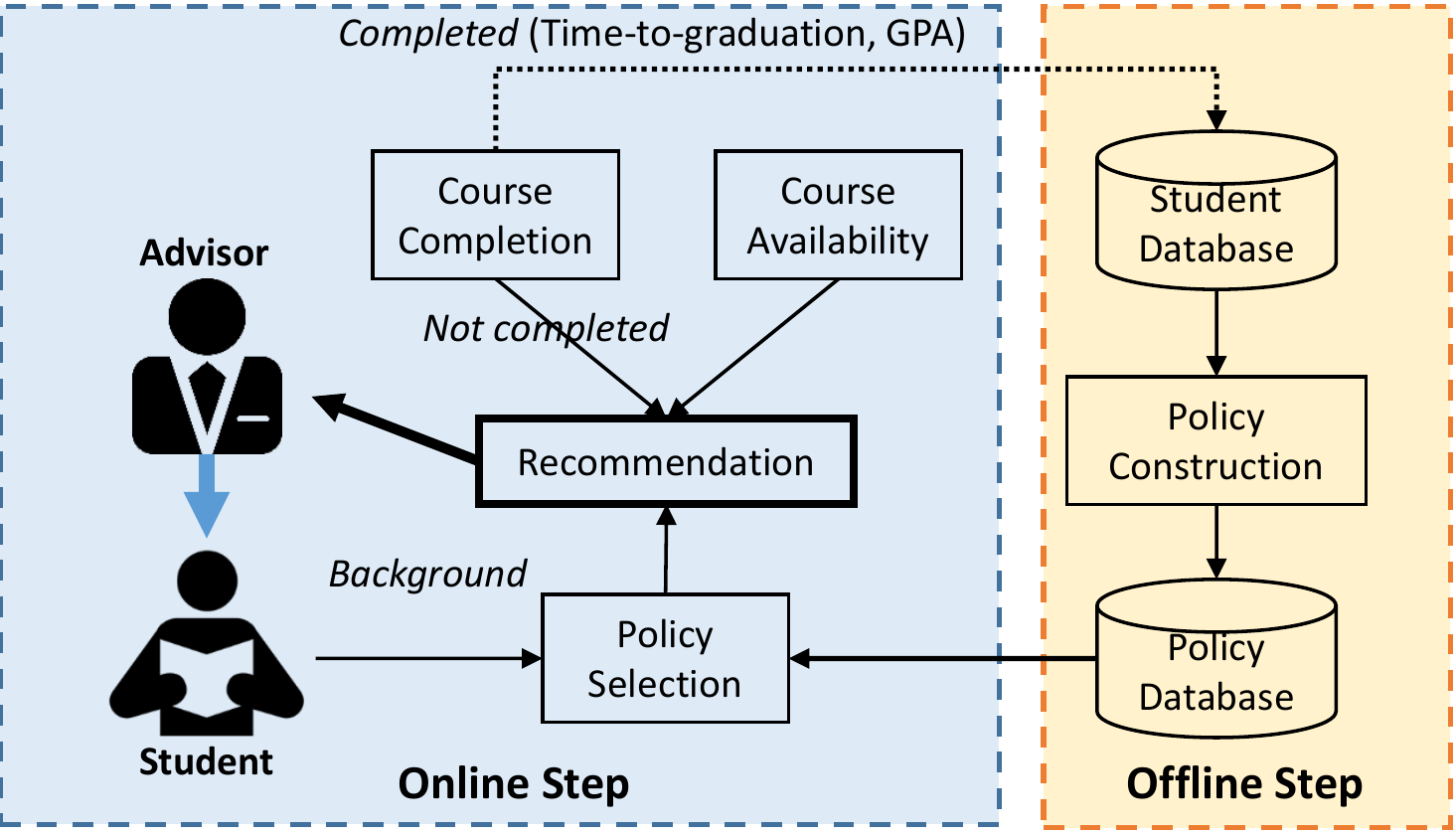}
	\caption{\small Illustration of course sequence recommendation.}
		\vspace{-0.1in}
	\label{fig:seq_system}
\end{figure}

In the school context, a specific challenge for developing a learning plan is the course sequence recommendation in degree programs \cite{xu2016personalized}. Recent studies show that the vast majority of college students in the United States do not complete college in the standard time. Moreover, today, compared to a decade ago, fewer college students graduate in a timely manner. While several factors contribute to students taking longer to graduate, such as credit losses in the transfer, uninformed choices due to the low advisor-student ratios, and poor preparation for college, the inability of students to attend the required courses is among the leading causes. If students select the courses myopically without a clear plan, they may end up in a situation where required subsequent courses are offered (much) later, thereby (significantly) prolonging the graduation time. Hence, to accelerate graduation, students shall essentially select courses in a foresighted way while taking the course sequences (shaped by courses being mandatory, elective, pre-requisite) into account. Moreover, it is vital to observe the timing in which the school offers various courses. More importantly, since the number and variety (in terms of backgrounds, knowledge, and goals) of students is expanding rapidly, the same learning path is unlikely to best serve all students. Therefore, it is crucially important to tailor course sequences to students. To this end, it is necessary to learn from the performance of previous students in various courses/sequences to adaptively recommend course sequences for the current students. Obviously, this depends on the student's background and his/her completion status of the program to maximize any of a variety of objectives including the time to graduation, grades, and the trade-off between the two. To make such plans, AI is a tool of great potential; however, designing AI technologies for personalized, foresighted, and adaptive course planning is challenging in several dimensions, as briefly described below.  
\begin{itemize}
\item First, course sequence recommendation requires dealing with a large decision space that grows combinatorially with the number of courses. 
\item Second, there is a great deal of flexibility in course sequence recommendation since multiple courses can be taken simultaneously while it is also subject to many constraints due to prerequisites and availability. 
\item Third, any static course sequence is sub-optimal since the knowledge, experience, and performance of a student develops and evolves in the process of learning. 
\item Last but not the least, students vary tremendously in backgrounds, knowledge, and goals. 
\end{itemize}
For example, in \cite{xu2016personalized}, we develop an automated course sequence recommendation system to address the aforementioned challenges. To reduce complexity and enabling tractable solutions, we solve the problem in two steps, as illustrated in \textbf{Figure \ref{fig:seq_system}}: (i) The first step corresponds to offline learning, in which a set of candidate recommendation policies are determined to minimize the expected time to graduation or the on-time graduation probability using an existing dataset of anonymized student records based on dynamic programming; (ii) The second step corresponds to online learning, in which for each new student, a suitable course sequence recommendation policy is selected depending on this student's background using the learned knowledge from the previous students. In other life-long learning contexts (e.g., the workplace), while similar challenges may still be present, new challenges are likely to emerge and hence, foresighted learning plans must be tailored to the specific context.

Recent research shows a significant gap between the lectures offered in schools and job requirements, especially in emerging disciplines like data science. Soft skills such as communication and teamwork are often even more important than typical technical skills \cite{katy}. Future research on life-long learning shall bridge this gap. Indeed, there is a systematic demand for the research community to identify and study the skills that significantly contribute to the professional perspectives instead of maximizing achievement in schools. The educators can take advantage of the findings to adjust school curricula and educational activities to better prepare students for the future. The centerpiece of possible approaches is to fuse a student's school records with future employment outcomes, possibly tracked over a long period, as well as other data sources such as course syllabi and job postings, to identify the crucial skills that extend from the education to profession. There is also a necessity for research in labor studies to conduct interviews with (i) employers to understand their requirements, (ii) job seekers to identify the skills they are keen to acquire, and (iii) training providers to clarify the skills that can be taught in a part-time or on-the-job way rather than through centralized educational programs, given workers' real-life constraints.
\section{Incentives and Motivation}
\label{sec:IM}
So far, one crucial aspect of personalized education has been largely left aside, namely motivation and incentive-design. This is unfortunate as these factors significantly contribute to the learners' perseverance and engagement, thereby the overall students' achievements. As such, they affect not only the individuals but also the entire society in terms of the efficiency of resource expenditure.\\
In educational sciences, motivation is regarded as a concept that involves several learning-related features such as initiation, goal-orientation, intensity, persistence, and quality of behavior \cite{Grove12:ISL}. Therefore, as \cite{Hartnett16:IMO} describes, motivation is an unobservable dynamic process that is difficult to be directly measured but it is inferable from the observations. Similar to other crucial factors of successful education such as talent and interest, motivation originates and is influenced by personal factors such as goals and beliefs. As such, it is reasonable to conclude that intelligent 'personalization' affects the motivation to a large extent.\\
Motivation can be intrinsic or triggered by external factors. As such, various features of personalized education, such as recommending a proper series of content or creating educational networks, might implicitly improve the learner's motivation by increasing the engagement level. Such efforts make the learning experience more pleasant, thereby improving the learner's satisfaction level. This is, however, insufficient. It is imperative to integrate direct motivating methods into personalized education and the learning platform. To this end, in the following, we first describe a few frameworks which can accommodate motivation and its relevant concepts appropriately. More can be found in \cite{Hartnett16:IMO}. 
\begin{itemize}
\item \textbf{Behavioral Economics:} Any personalized education platform shall be able to appropriately connect, interact, and interface with humans. Hence, the proper operation significantly depends on various features of the members of the target group that shapes their decision-making behavior. Indeed, a 'utility function' is the most seminal computational model for the interests of the learners. For a rational decision-maker, the utility function is conventionally increasing concave and to be maximized. However, humans often demonstrate unusual patterns in their utility functions and decision-making due to the following reasons: (i) Humans make mistakes, often due to inaccurate beliefs and imprecise predictions; (ii) Humans often act irrationally and based on heuristics; (iii) Humans think and act in different manners as a result of their unique background, including personality and experiences \cite{Maghsudi20:CMH}. Behavioral economics accommodate and formalizes such aspects; Therefore, one can take advantage of behavioral economics for efficient incentive design and motivation in learning platforms \cite{Levitt16:TBG}. 
\item \textbf{Self-Determination Theory:} This theory asserts that humans have an intrinsic urge to be self-autonomous, competent, and connected, concerning their environment \cite{Deci12:MPD}. While behavioral economics is appropriate to investigate the motivation resulted from external rewards, self-determination observes the motivation from an internal perspective. Indeed, any environment, including the learning platforms, that satisfies the aforementioned needs of human, awakes the intrinsic motivation, rendering external triggers rather unnecessary. As such, promoting intrinsic motivation is significantly more effective than extrinsic motivation as it is often associated with lower cost compared to material-rewarding and has a long-lasting effect \cite{Hartnett16:IMO}.
\item \textbf{Self-Efficacy Theory:} This concept corresponds to an individual's confidence in her capability of performing a specific task to be undertaken, for example, learning in an online learning platform or performing at a certain level \cite{Bandura97:SE}. Researchers show that humans constantly assess their self-efficacy, mainly based on the observed information from the environment and the past experiences \cite{Bandura97:SE}. Similar to the self-determination theory, self-efficacy considers the intrinsic motivation, implying that a feeling of efficacious triggers the internal motivation feeling in learners. Other relevant concepts include 'interest' and 'goal-orientation' \cite{Hartnett16:IMO}.  
\end{itemize}
The main challenge is to utilize AI and ML to motivate the learners of a personalized learning platform, based on the aforementioned theories that formalize and explain human behavior. To clarify this, consider the utility function of a learner in a personalized learning platform as an example \cite{Levitt16:TBG}. The function quantifies the learner's well-being while using the platform, and, consequently, her (future) engagement. Some learners exhibit hyperbolic preferences, overweighting the present so much that future rewards are largely ignored. Some learners show strong reactions even to non-monetary rewards. Some learners demonstrate reference-dependent preferences, implying that the utility is largely determined by its distance from a reference point, for example, a pre-defined goal or the average performance. By using ML and AI methods, the learning platform can take advantage of the available data and a learner's feedbacks to estimate the utility function of that learner, hence predicting her reaction to potential triggers of incentive and motivation . Consequently, the platform can adjust and allocate the reward among the learners efficiently and fairly. As another example, consider the self-determination theory. Based on this theory, a sophisticated personalized learning platform guarantees choice, connectedness, and the feeling of competence for the learner. To this end, the design of recommendation tools based on AI and ML methods should allow for enough alternatives, both at micro and macro levels, to ensure autonomy. Moreover, the suggested learning content should be based on the learner's feedback and the result's of accurate assessment, to avoid inducing a feeling of incompetence in the learner. Besides, promoting network formation or establishing a link between coherent learners and intensive interaction results in connectedness. This is also in accordance with the self-efficacy theory, in the sense that by providing an appropriate feedback and suitable side-information, the platform increase the positive belief of a learner in her ability to perform well on a learning platform. 
\section{Building Learning Networks}
\label{sec:BLE}
A potential negative effect of personalized education, especially in an online environment, is a loss of peer interactions and of the sense of community that is usually present in traditional classrooms. Fortunately, the rise of online social networks seems to facilitate interaction and networking between teachers and learners, also the co-production of content both within and outside the classroom. Learning applications and pedagogy can also be built based on online social networks to bridge formal and informal learning, also to promote peer interactions on both curricular and extra-curricular topics. Moreover, various education-related social networks have been created to facilitate collaboration, post/answer questions, and share resources; however, a formal method to build these learning networks and a deep understanding of their effectiveness are absent.

The core of learning networks is peer interaction, which has important implications for personalized education when teaching resources are limited. For example, peer review serves as an effective and scalable method for assessment and evaluation when the number of students enrolled in a course far exceeds the number of teaching assistants. However, effective peer review in learning networks poses new challenges \cite{xiao2018incentive}. On the one hand, the peer reviewers have different intrinsic capabilities, which are often unknown. On the other hand, the peer reviewers can choose to exert different levels of effort (e.g., time and energy spent in reviewing), which is unobservable. Identifying unknown intrinsic capabilities corresponds to the \textit{adverse selection problem} in the game theory. A natural candidate for solving this problem is to use matching mechanisms, i.e., assign reviewers to students. Existing works on matching mechanisms focus on one-shot peer interactions and design one-shot matching rules. However, their assumption does not hold in peer review systems, where the review quality depends crucially on the reviewers' effort. Motivating reviewers to exert high effort corresponds to the \textit{moral hazard problem} in game theory. One way to address this problem is to use social norms, where each peer reviewer is assigned with a rating that summarizes her past behavior and recommended a 'norm' that rewards reviewer with good ratings and punishes those with bad ratings. However, existing works on social norms assume that peer reviewers are homogeneous. This assumption does not hold in peer review systems because different reviewers have different intrinsic capabilities. Because a peer reviewer's ultimate review quality is determined by her intrinsic capabilities and effort, designing effective peer review systems in learning networks becomes significantly more challenging due to the presence of both adverse selection and moral hazard. Therefore, new peer review system designs shall simultaneously solve both problems so that peer reviewers find it in their self-interest to exert high effort and receive ratings that truly reflect their capabilities. 

Another primary function of learning networks is to foster learning content co-production and sharing. Building such learning networks is vastly different from building traditional networks such as computer networks and transportation networks, as in learning networks, individual learners create and maintain the links. Because links permit the acquisition and dissemination of learning content, it is theoretically intriguing and practically valuable to have a deeper understanding of the networks that are more likely to be formed by self-interested learners. Game theory is a useful tool to formulate and understand the strategic behavior of learners. The formulation must capture the heterogeneity of learners in terms of goals, capabilities, costs, and self-interest nature \cite{Maghsudi19:DTM}; That is, each learner intends to maximize its benefit from content co-production and sharing minus whatever cost it pays to establish links. Our prior work \cite{zhang2013strategic} studies the endogenous formation of networks by strategic, self-interested agents who benefit from producing and disseminating information. The results showed that the typical network structure that emerges in equilibrium displays a core-periphery structure, with a smaller number of agents at the core of the network and a larger number of agents at the periphery of the network. Furthermore, we established that the typical networks that emerge are minimally connected and have short network diameters, which are independent of the size of the network. In other words, theoretical results show that small diameters tend to make information dissemination efficient and minimal connectivity leads to minimizing the total cost of constructing the network. These results are consistent with the outcome of numerous empirical investigations. Such theoretical analysis and tools are essential to guide building learning networks. Also, based on this analysis, one can create protocols to motivate selfish learners to take actions that promote the system-wide utility.  

Future research in learning networks hinges on understanding the knowledge flow between students via peer interaction. Such an understanding enables educators to effectively modulate peer interactions and to encourage the interactions that promote peer learning. Peer learning is especially valuable as education extends to more diverse settings, such as remote online learning during the COVID-19 pandemic. In such settings, it is hard for instructors to moderate learning activities remotely; hence, peer learning through online course discussion forums becomes essential. Therefore, it is vital to understand the interaction tendencies and students' behaviors in these discussion forums \cite{moocforum}, understanding the flow of knowledge by combining discussion forum activities with grades, identifying the factors that enhance knowledge flow, and designing automated strategies to moderate student activities when necessary. 
\section{Diversity, Fairness, and Biases}
\label{sec:DFB}
Experimental studies show that AI-driven personalization such as student assessment, feedback, and content recommendation improve the overall learning outcomes; nonetheless, certain student subgroups may benefit more than other subgroups due to the biases that exist in the training data \cite{Reich:Good:2017}. This imbalance jeopardizes students that are already under-served in particular since they often have less access to advanced, digitized educational systems and are less frequently represented in datasets collected by these systems \cite{Doroudi:Fairer:2019}. Therefore, it is essential to develop AI tools that promote fairness among learners with different backgrounds, thereby making education more inclusive for the next generations. 

To mitigate biases and to promote fairness and equity in AIs, currently, the researchers pay significant attention to developing approaches that promote fairness, primarily in the context of predictive algorithms. 
\begin{itemize}
\item The first major research problem studied is how to properly define fairness; see \cite{Gajane:Formalizing:2017} and references therein for an overview. Many definitions of fairness exist, including individual fairness that requires that users with similar feature values be treated similarly, parity in the predicted probability of each outcome across user groups (drawn using sensitive attributes), parity in the predicted probability of each outcome given actual comes and regardless of sensitive attributes, and counterfactual fairness, which requires that the predicted outcome for each user remain mostly unchanged if the sensitive attribute changes.
\item The second major research problem is to develop methods that enforce fairness in predictive algorithms. Existing approaches include preprocessing the data to select only the fair features as input to algorithms, and post-processing the output of algorithms to balance across user groups. The most promising approach is to impose regularizers and constraints while training predictive algorithms. These methods result in better fairness at the expense of sacrificing some classification accuracy; however, they are empirically shown to obtain better tradeoffs between fairness and accuracy than other fairness-promoting methods.
\end{itemize}
Promoting fairness and equity is a crucial necessity of education that requires a comprehensive approach to be fulfilled: We need to not only design fair personalization algorithms but also develop systematic principles and guidelines for their application in practice; In other words, we need a set of tools to regulate the use of AI algorithms.\\
Finally, despite great promises, AI-driven personalization in education can also bring risks that have to be closely monitored and controlled. Recently, there have been calls for a food and drug administration (FDA)-type framework for other AI applications such as facial recognition \cite{erik}. It is essential to establish a similar ecosystem in education with a set of legislation and regulations around issues of data ownage, sharing, continuous performance monitoring, and validation, to regulate every step of the process, from ensuring the diversity and quality of the collected data, developing algorithms with performance guarantees across different educational settings, to identifying misuse and implementing a fail-safe mechanism. 
\section{COVID-19 and AI-Enabled Personalized Education}
\label{sec:Covid}
Among its several other adverse effects, the COVID-19 pandemic has disrupted or interrupted the functionality of conventional education systems around the globe. Not surprisingly, the students perceive the bitterness of this adverse effect at various degrees, depending on several factors such as country/region, family status, and individual characteristics. The complications vary over a large spectrum and include reduced learning ability, depression, loss of concentration, and a decline in physical fitness. The complications mainly arise due to spending less- or no time at school, where the students receive educational materials and support in learning, interact with their peers and teachers, develop incentives, and are evaluated. Besides, many students cannot take full advantage of replacements such as online materials, e.g., in the absence of an appropriate technological device or reliable internet connection, or suitable learning ambient at home. As a consequence of its vitality, the impacts of COVID-19 on education has attracted a great deal of attention. For example, in \cite{Garcia20:CVP}, the authors describe the influence of pandemic-triggered growth in online learning on student's performance and equity. Another example is \cite{Pietro20:LIC}, which also provides suggestions for policy-makers to compensate for the pandemic's negative educational consequences. Moreover, some research works such as \cite{Almarzooq20:ADT} study domain-specific educational effects of the pandemic and evaluate the available solutions. 
 
Personalized- and distance education have already been trending in the past decade; Still, COVID-19 has urged both public and private sectors to rapidly increase the investigations into research and development in this area to earn individual- and/or social profit. For example, the pandemic has boosted the usage of online learning tools for signal processing education, especially at the undergraduate levels. These include, for example, web-based laboratories for digital signal processing \cite{Dixit18:OML}, and online machine learning education modules \cite{Shanthamallu19:IML}. While it is essential to carefully study such a tremendous push towards revolutionizing education from several perspectives, in the scope of our paper, we confine our attention to the role and influence of AI and ML.

As described previously, AI and ML have a great potential to enhance online education in different ways, e.g., through improving the quality of learning materials, enabling fairness and diversity, generating proper tests, and allowing to build knowledge networks. That side is a universal aspect of applying AI and ML methods in distance and asynchronous education regardless of the current pandemic; nevertheless, such methods can additionally assist in accelerating rebuilding the education systems and in mitigating the pandemic's detrimental effects. For example, by using ML methods on the available data, policy-makers can classify the students based on the exposure to the educational effects of a pandemic; using such classification, one can allocate resources efficiently while satisfying fairness constraints. As another example, b taking advantage of ML methods, one can optimize the school closure plan based on different features such as neighborhood, size, grade, and the like.
\section{Summary and Conclusion}
\label{sec:Conc}
Enabling 'personalized education' is one of the most precious merits of AI concerning education. This paradigm significantly improves the quality of education in several dimensions by adapting to the distinct characteristics and expectations of each learner such as personality, talent, objectives, and background. Besides, online education is of the utmost value under abnormal circumstances such as the COVID-19 outbreak or natural disasters. Indeed, conventional education requires significantly more resources than the online format concerning educational space, scheduling, and human resources, which makes it prone to failure with even a small shift in conditions. As such, emerging alternatives are inevitable. Despite having the potential of a revolutionary transformation from traditional education to modern concepts, personalized education is associated with several challenges. We discussed such challenges, provided a brief overview of the state-of-the-art research, and proposed some solutions. \textbf{Table \ref{Tb:Challenges}} summarizes some of the future research directions.
\begin{table}[h!]
\scriptsize
\centering
\caption{Some Research Directions for AI-based Personalized Education}
\begin{tabular}{|c||c||c|}
\hline 
\textbf{Challenge} & \textbf{Description} & \textbf{Some References}\\ \hline \hline

\shortstack{Content Production/Recommendation\\ \textcolor{white}{.}}&
\shortstack{Personalized and profession-oriented production,\\ recommendation, and maintenance of contents}&
\shortstack{\cite{mozpash}, \cite{pnasrep}, \cite{mozeraz}, \cite{banditslan}
\\ \textcolor{white}{.}}
\\\hline
\shortstack{Evaluation and Assessment\\ \textcolor{white}{.}}& 
\shortstack{Performance comparison in personalized education,\\ testing without information loss, accreditation}& 
\shortstack{\cite{akt}, \cite{howdeep}, \cite{qgen}, \cite{lafferty}, \cite{neilcrowd}\\ \textcolor{white}{.}}
\\ \hline
\shortstack{Lifelong Learning\\ \textcolor{white}{.}}& 
\shortstack{Continuous education and additional qualification\\ for improvement and pivots in profession}& 
\shortstack{\cite{xu2016personalized}\\ \textcolor{white}{.}}
\\ \hline
\shortstack{Incentives\\ \textcolor{white}{.}}& 
\shortstack{Internal and external motivation for learning,\\ gamification, rewarding, inducing confidence}& 
\shortstack{\cite{Hartnett16:IMO}, \cite{Levitt16:TBG} \\ \textcolor{white}{.}}
\\ \hline
\shortstack{Networking and Interaction\\ \textcolor{white}{.}}& 
\shortstack{Inducing learning networks, forming coalitions\\ for efficient learning, imitating teacher feedback}&
\shortstack{\cite{xiao2018incentive}, \cite{moocforum}\\ \textcolor{white}{.}}
\\ \hline
\shortstack{Diversity and Fairness\\ \textcolor{white}{.}}& 
\shortstack{Equal access to quality online education,\\
avoiding biases in platform development}& \shortstack{\cite{Gajane:Formalizing:2017}, \cite{Reich:Good:2017}\\ \textcolor{white}{.}}
\\ \hline
\end{tabular}
\label{Tb:Challenges}
\end{table}
\bibliographystyle{IEEEbib}
\bibliography{main}

\begin{thebibliography}{10}

\bibitem{weixu10:DLF}
Y.~Zhong, C.~Jiang, W.~Xu, and J.~Li,
\newblock ``Discourse level factors for sentence deletion in text
  simplification,''
\newblock in {\em Proc. AAAI Conference on Artificial Intelligence}, Feb. 2020.

\bibitem{sachan}
M.~Sachan, A.~Dubey, E.~H. Hovy, T.~M. Mitchell, D.~Roth, and E.~P. Xing,
\newblock ``Discourse ineffectiveness of crowd-sourcing on-demand tutoring from
  teachers in online learning platforms multimedia: {A} case study in
  extracting geometry knowledge from textbooks,''
\newblock {\em Computational Linguistics}, vol. 45, no. 4, pp. 627--665, 2020.

\bibitem{Manickam17:CMAB}
I.~{Manickam}, A.~S. {Lan}, and R.~G. {Baraniuk},
\newblock ``Contextual multi-armed bandit algorithms for personalized learning
  action selection,''
\newblock in {\em 2017 IEEE International Conference on Acoustics, Speech and
  Signal Processing (ICASSP)}, 2017, pp. 6344--6348.

\bibitem{Ghauth10:LMR}
K.I. Ghauth and N.A. Abdullah,
\newblock ``Learning materials recommendation using good learners’ ratings
  and content-based filtering,''
\newblock {\em Education Tech Research and Development}, , no. 58, pp.
  711–727, 2010.

\bibitem{Magno09:DDC}
C.~Magno,
\newblock ``Demonstrating the difference between classical test theory and item
  response theory using derived test data,''
\newblock {\em CSN: General Cognitive Social Science (Topic)}, vol. 1, 06 2009.

\bibitem{pardoskt}
Z.~Pardos and N.~Heffernan,
\newblock ``Modeling individualization in a {B}ayesian networks implementation
  of knowledge tracing,''
\newblock in {\em Proc. International Conference on User Modeling, Adaptation,
  and Personalization}, June 2010, pp. 255--266.

\bibitem{Sharples00:DPM}
M.~Sharples,
\newblock ``The design of personal mobile technologies for lifelong learning,''
\newblock {\em Computers and Education}, vol. 34, no. 3, pp. 177 -- 193, 2000.

\bibitem{Parisi19:CLL}
G.~I. Parisi, R.~Kemker, J.~L. Part, C.~Kanan, and S.~Wermter,
\newblock ``Continual lifelong learning with neural networks: A review,''
\newblock {\em Neural Networks}, vol. 113, pp. 54 -- 71, 2019.

\bibitem{Grove12:ISL}
W.~Grove and L.~Hadsell,
\newblock {\em Open Learning Environments}, chapter Incentives and Student
  Learning, pp. 1511--1517,
\newblock Springer, 01 2012.

\bibitem{Buckley14:ILE}
P.~Buckley and E.~Doyle,
\newblock ``Gamification and student motivation,''
\newblock {\em Interactive Learning Environments}, vol. 24, no. 6, pp.
  1162--1175, 2016.

\bibitem{Chen10:AAA}
K.-C. Chen, S.-J. Jang, and R.~M. Branch,
\newblock ``Autonomy, affiliation, and ability: Relative salience of factors
  that influence online learner motivation and learning outcomes,''
\newblock {\em Knowledge Management and E-Learning}, vol. 2, no. 1, pp.
  1162--1175, 2010.

\bibitem{Rovai00:BSC}
A.~P. Rovai,
\newblock ``Building and sustaining community in asynchronous learning
  networks,''
\newblock {\em The Internet and Higher Education}, vol. 3, no. 4, pp. 285 --
  297, 2000.

\bibitem{Vesely07:KEB}
P.~Vesely, L.~Bloom, and J.~Sherlock,
\newblock ``Key elements of building online community: Comparing faculty and
  student perceptions,''
\newblock {\em MERLOT Journal of Online Learning and Teaching}, vol. 3, 01
  2007.

\bibitem{Gardner:Evaluating:2019}
J.~Gardner, C.~Brooks, and R.~Baker,
\newblock ``Evaluating the fairness of predictive student models through
  slicing analysis,''
\newblock in {\em Proceedings of the 9th International Conference on Learning
  Analytics \& Knowledge}, 2019, pp. 225--234.

\bibitem{Yao:Beyond:2017}
S.~Yao and B.~Huang,
\newblock ``Beyond parity: Fairness objectives for collaborative filtering,''
\newblock in {\em Advances in Neural Information Processing Systems}, 2017, pp.
  2921--2930.

\bibitem{qgen}
Z.~Wang, A.~S. Lan, W.~Nie, A.~E. Waters, P.~J. Grimaldi, and R.~G. Baraniuk,
\newblock ``{QG}-net: {A} data-driven question generation model for educational
  content,''
\newblock in {\em Proc. ACM Conference on Learning at Scale}, June 2018, pp.
  1--10.

\bibitem{lafferty}
M.~Yasunaga and J.~D Lafferty,
\newblock ``{TopicEq}: A joint topic and mathematical equation model for
  scientific texts,''
\newblock in {\em Proc. AAAI Conference on Artificial Intelligence}, 2019, pp.
  7394--7401.

\bibitem{neilcrowd}
P.~Thanaporn and N.~T. Heffernan,
\newblock ``Effectiveness of crowd-sourcing on-demand tutoring from teachers in
  online learning platforms,''
\newblock in {\em Proc. ACM Conference on Learning at Scale}, Aug. 2020, pp.
  1--10.

\bibitem{Zylich:Exploring:2020}
B.~Zylich, A.~Viola, B.~Toggerson, L.~Al-Hariri, and A.~S. Lan,
\newblock ``Exploring automated question answering methods for teaching
  assistance,''
\newblock in {\em Proc. International Conference on Artificial Intelligence in
  Education (AIED)}, July 2020.

\bibitem{banditslan}
A.~S. Lan and R.~G. Baraniuk,
\newblock ``A contextual bandits framework for personalized learning action
  selection,''
\newblock in {\em Proc. International Conference on Educational Data Mining},
  June 2016, pp. 424--429.

\bibitem{xu2016personalized}
J.~Xu, T.~Xing, and M.~Van Der~Schaar,
\newblock ``Personalized course sequence recommendations,''
\newblock {\em IEEE Transactions on Signal Processing}, vol. 64, no. 20, pp.
  5340--5352, 2016.

\bibitem{mozeraz}
M.~C. Mozer,
\newblock ``Bayesian optimization for exploratory experimentation in the
  behavioral sciences,'' online:
  {https://www.cs.colorado.edu/~mozer/index.php?dir=/Research/Projects/Bayesian
  optimization/}.

\bibitem{pnasrep}
B.~Tabibian, U.~Upadhyay, A.~De, A.~Zarezade, B.~Sch{\"o}lkopf, and
  M.~Gomez-Rodriguez,
\newblock ``Enhancing human learning via spaced repetition optimization,''
\newblock {\em Proceedings of the National Academy of Sciences}, vol. 116, no.
  10, pp. 3988--3993, 2019.

\bibitem{Tekin15:OLP}
C.~{Tekin}, J.~{Braun}, and M.~{van der Schaar},
\newblock ``etutor: Online learning for personalized education,''
\newblock in {\em 2015 IEEE International Conference on Acoustics, Speech and
  Signal Processing (ICASSP)}, 2015, pp. 5545--5549.

\bibitem{lordirt}
F.~Lord,
\newblock {\em Applications of Item Response Theory to Practical Testing
  Problems},
\newblock Erlbaum Associates, 1980.

\bibitem{mirt}
M.~D. Reckase,
\newblock {\em Multidimensional Item Response Theory},
\newblock Springer, 2009.

\bibitem{bkt}
M.~Yudelson, K.~Koedinger, and G.~Gordon,
\newblock ``Individualized {B}ayesian knowledge tracing models,''
\newblock in {\em Proc. International Conference on Artificial Intelligence in
  Education}, July 2013, pp. 171--180.

\bibitem{pfa}
P.~Pavlik~Jr, H.~Cen, and K.~Koedinger,
\newblock ``Performance factors analysis--{A} new alternative to knowledge
  tracing,''
\newblock in {\em Proc. International Conference on Artificial Intelligence in
  Education}, June 2009, pp. 531--538.

\bibitem{dkt}
C.~Piech, J.~Bassen, J.~Huang, S.~Ganguli, M.~Sahami, L.~J. Guibas, and
  J.~Sohl-Dickstein,
\newblock ``Deep knowledge tracing,''
\newblock in {\em Proc. Conference on Advances in Neural Information Processing
  Systems}, Dec. 2015, pp. 505--513.

\bibitem{howdeep}
M.~Khajah, R.~Lindsey, and M.~Mozer,
\newblock ``How deep is knowledge tracing?,''
\newblock in {\em Proc. International Conference on Educational Data Mining},
  July 2016, pp. 94--101.

\bibitem{akt}
A.~Ghosh, T.~Heffernan, and A.~S. Lan,
\newblock ``Context-aware attentive knowledge tracing,''
\newblock in {\em Proc. ACM SIGKDD International Conference on Knowledge
  Discovery and Data Mining}, Aug. 2020.

\bibitem{katy}
K.~B{\"o}rner, O.~Scrivner, M.~Gallant, S.~Ma, X.~Liu, K.~Chewning, L.~Wu, and
  J.~A. Evans,
\newblock ``Skill discrepancies between research, education, and jobs reveal
  the critical need to supply soft skills for the data economy,''
\newblock {\em Proceedings of the National Academy of Sciences}, vol. 115, no.
  50, pp. 12630--12637, 2018.

\bibitem{Hartnett16:IMO}
M.~Hartnett,
\newblock {\em Motivation in Online Education}, chapter The Importance of
  Motivation in Online Learning, pp. 5--32,
\newblock Springer, 01 2016.

\bibitem{Maghsudi20:CMH}
S.~{Maghsudi} and M.~{Davy},
\newblock ``Computational models of human decision-making with application to
  the internet of everything,''
\newblock {\em IEEE Wireless Communications}, pp. 1--8, 2020.

\bibitem{Levitt16:TBG}
S.~D. Levitt, J.~A. List, S.~Neckermann, and S.~Sadoff,
\newblock ``The behavioralist goes to school: Leveraging behavioral economics
  to improve educational performance,''
\newblock {\em American Economic Journal: Economic Policy}, vol. 8, no. 4, pp.
  183--219, November 2016.

\bibitem{Deci12:MPD}
E.~Deci and R.~Ryan,
\newblock ``Motivation, personality, and development within embedded social
  contexts: An overview of self-determination theory,''
\newblock {\em The Oxford Handbook of Human Motivation}, 01 2012.

\bibitem{Bandura97:SE}
A.~Bandura,
\newblock {\em Self-efficacy: The exercise of control},
\newblock Freeman, 1997.

\bibitem{xiao2018incentive}
Y.~Xiao, F.~D{\"o}rfler, and M.~Van Der~Schaar,
\newblock ``Incentive design in peer review: Rating and repeated endogenous
  matching,''
\newblock {\em IEEE Transactions on Network Science and Engineering}, vol. 6,
  no. 4, pp. 898--908, 2018.

\bibitem{Maghsudi19:DTM}
S.~{Maghsudi} and M.~{van der Schaar},
\newblock ``Distributed task management in cyber-physical systems: How to
  cooperate under uncertainty?,''
\newblock {\em IEEE Transactions on Cognitive Communications and Networking},
  vol. 5, no. 1, pp. 165--180, 2019.

\bibitem{zhang2013strategic}
Y.~Zhang and M.~van~der Schaar,
\newblock ``Strategic networks: Information dissemination and link formation
  among self-interested agents,''
\newblock {\em IEEE Journal on Selected Areas in Communications}, vol. 31, no.
  6, pp. 1115--1123, 2013.

\bibitem{moocforum}
A.~S. Lan, J.~Spencer, Z.~Chen, C.~Brinton, and M.~Chiang,
\newblock ``Personalized thread recommendation for {MOOC} discussion forums,''
\newblock in {\em Proc. European Conf. Mach. Learn. and Principle Knowl.
  Discov. Databases}, Sep. 2018.

\bibitem{Reich:Good:2017}
J.~Reich and M.~Ito,
\newblock ``From good intentions to real outcomes: Equity by design in learning
  technologies,''
\newblock {\em Irvine, CA: Digital Media and Learning Research Hub}, 2017.

\bibitem{Doroudi:Fairer:2019}
S.~Doroudi and E.~Brunskill,
\newblock ``Fairer but not fair enough on the equitability of knowledge
  tracing,''
\newblock in {\em Proceedings of the 9th International Conference on Learning
  Analytics \& Knowledge}. ACM, 2019, pp. 335--339.

\bibitem{Gajane:Formalizing:2017}
P.~Gajane and M.~Pechenizkiy,
\newblock ``On formalizing fairness in prediction with machine learning,''
\newblock {\em arXiv preprint arXiv:1710.03184}, 2017.

\bibitem{erik}
E.~Learned-Miller, V.~Ordonez, J.~Morgenstern, and J.~Buolamwini,
\newblock ``Facial recognition technologies in the wild: {A} call for a federal
  office,'' online: {https://people.cs.umass.edu/~elm/papers/FRTintheWild.pdf}.

\bibitem{Garcia20:CVP}
E.~Garcia and E.~Weiss,
\newblock ``Covid-19 and student performance, equity, and u.s. education
  policy,'' 2020.

\bibitem{Pietro20:LIC}
G.~Di Pietro, F.~Biagi, P.~D. Mota~Da Costa, Z.~Karpinski, and J.~Mazza,
\newblock ``The likely impact of covid-19 on education: Reflections based on
  the existing literature and recent international datasets,''
\newblock {\em Publications Office of the European Union (online)}, 2020.

\bibitem{Almarzooq20:ADT}
Z.~I. Almarzooq, M.~Lopes, and A.~Kochar,
\newblock ``Virtual learning during the {COVID-19} pandemic: A disruptive
  technology in graduate medical education,''
\newblock {\em Journal of the American College of Cardiology}, vol. 75, no. 20,
  pp. 2635--2638, 2020.

\bibitem{Dixit18:OML}
A.~{Dixit}, U.~S. {Shanthamallu}, A.~{Spanias}, V.~{Berisha}, and M.~{Banavar},
\newblock ``Online machine learning experiments in {HTML5},''
\newblock in {\em IEEE Frontiers in Education Conference}, 2018, pp. 1--5.

\bibitem{Shanthamallu19:IML}
U.~S. {Shanthamallu}, S.~{Rao}, A.~{Dixit}, V.~S. {Narayanaswamy}, J.~{Fan},
  and A.~{Spanias},
\newblock ``Introducing machine learning in undergraduate dsp classes,''
\newblock in {\em International Conference on Acoustics, Speech and Signal
  Processing}, 2019, pp. 7655--7659.

\bibitem{mozpash}
N.~J. Cepeda, N.~Coburn, D.~Rohrer, J.~T. Wixted, M.~C. Mozer, and H.~Pashler,
\newblock ``Optimizing distributed practice: {T}heoretical analysis and
  practical implications,''
\newblock {\em Experimental psychology}, vol. 56, no. 4, pp. 236--246, 2009.

\end{thebibliography}
\end{document}